\documentclass{JINST}

\usepackage{subfigure}

\usepackage{lineno}
\usepackage{url}

\title{The Electronics and Data Acquisition System for the PandaX-I Dark Matter Experiment}

\author{Xiangxiang Ren$^{a,b}$\thanks{Corresponding author, renxx@sjtu.edu.cn},
Xun Chen$^{b}$,
Xiangdong Ji$^{b,c,d}$,
Shaoli Li$^{b}$,
Siao Lei$^{b}$,
Jianglai Liu$^{b}$,
Meng Wang$^a$\thanks{Corresponding author, mwang@sdu.edu.cn},
Mengjiao Xiao$^{b}$,
Pengwei Xie$^{b}$,
Binbin Yan$^a$\\
\llap{$^a$} School of Physics and Key Laboratory of Particle Physics and Particle Irradiation (MOE), Shandong University, Jinan 250100, China,\\
\llap{$^b$}INPAC and Department of Physics and Astronomy,
Shanghai Jiao Tong University, \\
Shanghai Laboratory for Particle Physics and Cosmology, Shanghai, 200240, China,\\
\llap{$^c$}Center for High-Energy Physics, Peking University, Beijing, 100080, China\\
\llap{$^d$}Department of Physics, University of Maryland, College Park, MD, 20742, USA\\
} \abstract{ We describe the electronics and data acquisition system
  used in the first phase of the PandaX experiment -- a 120 kg
  dual-phase liquid xenon dark matter direct detection experiment in
  the China Jin-Ping Underground Laboratory.  This system utilized 180
  channels of commercial flash ADC waveform digitizers.  During the entire experimental run, the system
  has achieved low trigger threshold ($<$1 keV electron-equivalent energy)
  and low deadtime data acquisition.
}

\begin{document}

\section{Introduction}
The PandaX project consists of a series of experiments at the
China Jin-Ping Underground Laboratory, utilizing xenon-based
time projection chambers (TPC). The first phase experiment,
PandaX-I~\cite{ref:pandax_tdr, ref:pandax_paper1, ref:pandaxI_final},
with a 120-kg dual-phase xenon dark matter detector, completed its
data taking in October 2014. The TPC of PandaX-I had a cylindrical shape with
60 cm diameter and 15 cm height, confined by the cathode plane located at
the bottom in the liquid xenon, and a gate and anode planes located 4 mm
below and above the liquid level, respectively.
Particle interactions in the xenon produce prompt scintillation photons (S1).
Some of the electrons from ionization produced in the same interaction drift vertically
upward between the cathode and gate,
and get extracted by a stronger field between the gate
and the anode into the gas, producing proportional scintillation photons (S2).
The background gammas produce electron recoil events,
while the dark matter signals interact with xenon atomic nucleus
producing nuclear recoil events. The ratio of S2 to S1 is less for the latter
compared to the former due to different ionization capability.
This ratio is therefore used as an effective background
discrimination parameter.
Two arrays of photomultiplier
tubes (PMTs) consisting of 143 1-inch R8520-406 tubes and
37 3-inch R11410-MOD tubes from Hamamatsu~\cite{ref:hamamatsu}
were located at the top and bottom, respectively, looking into the TPC collecting S1s and
S2s. The time difference between S1 and S2 signals gives
the vertical position of the interaction, and the charge pattern of S2 signals on the PMT arrays
encodes the horizontal position.

Being an underground experiment for very rare event search, the design
philosophy of the electronics and data acquisition (DAQ) system of PandaX-I was to
maximize the data collection under the lowest possible threshold, and leave the
``data mining'' work to the offline analysis.
The system digitizes the waveforms from the PMTs, generates triggers on
candidate signals, and saves the data to disks. The remainder of this paper is
organized as follows. In Sec.~\ref{sec:requirements}, the requirements of the
system are presented, followed by the design details of the electronics in
Sec.~\ref{sec:elec} and DAQ in Sec.~\ref{sec:daq}.
We summarize the performance of the system
during the experiment in Sec.~\ref{sec:performance} before the conclusion.

\section{System Requirements}
\label{sec:requirements}
The typical dark matter recoil signals deposit small amounts of energy
from sub-keV to a few tens of keV. For PandaX-I, this translates into a typical S1
from 1 to 100 photoelectrons (PE) with a time range of 100 ns, combining all PMTs.
The intensity of the corresponding S2 signal is typically one hundred times that
of S1 (100 to 10000 PE)
with a time span of  $\sim$2~$\mu$s determined by the properties of the gas and the
gap between the gate and anode. There are two general requirements for
the PandaX-I electronics.
\begin{itemize}
\item To achieve a low energy threshold of $<$1 keV$_{ee}$ electron equivalent energy
(relating to the true nuclear recoil energy by the so-called Lindhard factor).
\item To digitize and read out the waveform faithfully from each PMTs for offline analysis.
\end{itemize}

\begin{table}[!hbp]
\centering
\begin{tabular}{cc}
\hline
Trigger threshold & $<$1 keV$_{ee}$ \\
S1 signal & [1, 100] PE in 100 ns \\
S2 signal & [100, 10000] PE in 2 $\mu$s \\
Readout window & $100~\mu$s pre- and post-trigger width\\
Digitizer ADC resolution & 14 bit, 2 V range \\
Digitizer sampling rate & 100 MS/s \\
Maximum rate & 30 Hz (source calibration), 10 Hz (dark matter), 100 Hz (LED) \\
Number of channels & 180 \\
Maximum data bandwidth & 70 MB/s with baseline suppression\\
\hline
\end{tabular}
\caption{Specific electronics and DAQ requirements for PandaX-I.}
\label{tab:req}
\end{table}

These two general requirements are shown more specifically in Table~\ref{tab:req}.
For PandaX-I, the maximum electron drift length is
15 cm, corresponding to an 88 $\mu$s maximum separation between the S1 and S2
with a drift speed of $\sim$1.7~mm/$\mu$s. To allow the trigger to be generated either
by S1 or S2, the readout window should have an equal pre- and post-trigger width of
at least 88~$\mu$s.
To be able to cleanly monitor random light generation in the detector,
the readout window was selected to be $\pm$100~$\mu$s before and after the trigger.
The single photoelectron (SPE) signals, assuming an overall gain of 2$\times$10$^7$,
are about 10 mV in amplitude with a full-width-half-maximum of about 20~ns and a full range of about 100~ns (see later Fig.~\ref{fig:waveform}),
resulting in an approximate 10 mV$\times$20 ns conversion factor from pulse area to a SPE.
According to a simulation including digitization and bandwidth effects,
in order to achieve a $<$5$\%$ bias to the charge of a single photoelectron,
a 100 MS/s sampling rate and a 14 bit ADC (with full range $\sim$2 V) is required.

The requirement on data bandwidth depends on the types of data runs in PandaX-I.
The first type is the low occupancy LED run to calibrate the gains of the PMTs,
in which the fast light pulses from external LEDs were injected into the detector
with the system triggered by the LED driving pulses ($\sim$ 100 Hz).
Such a type of run can have a short
readout window ($\sim1~\mu$s) and therefore has a small event size.
The second type is the calibration run with gammas and neutrons, which
is required to achieve a rate of $\sim$30 Hz in order to collect sufficient
calibration data within a few days.
The third type is the normal dark matter search run, in which the
trigger rate is expected to be $<$10 Hz based on the Monte Carlo simulation.
Clearly the latter two types are required to have a long readout window
of $\sim$200$\mu$s,
and the source calibration runs dictate the data bandwidth requirement.
Taking the
14 bit ADC and 100 MS/s sampling rate, the estimated bandwidth is 210 MB/s
if no data reduction is used on the digitizer data. In reality, the baseline data can
be suppressed (see Sec.~\ref{sec:zle}), leading to at least a factor 3 reduction in
data rate (70~MB/s).


\section{System Design}
\label{sec:elec}
\subsection{Electronics System}
\begin{figure}[!htbp]
\centering
\includegraphics[width=1\textwidth]{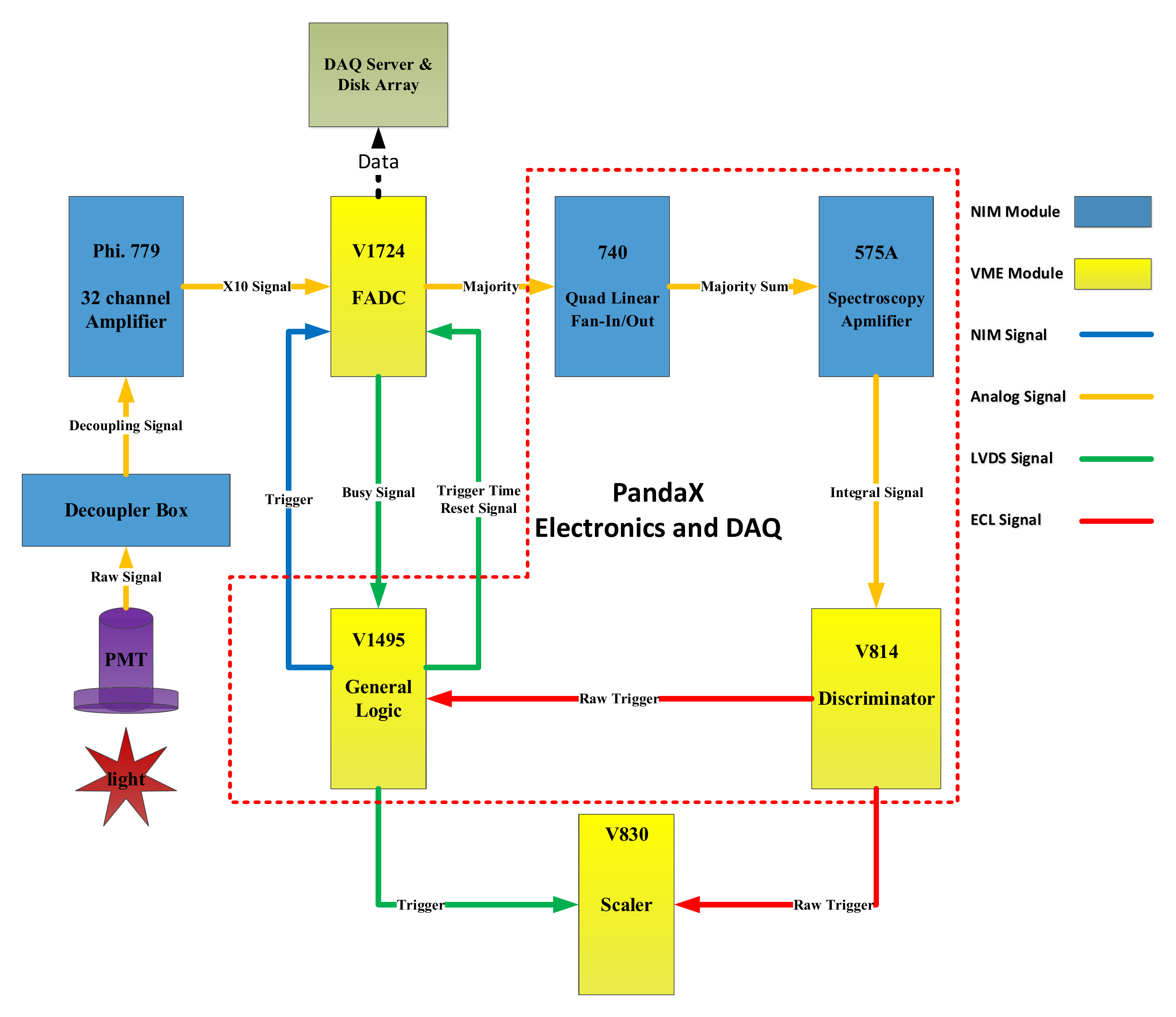}
\caption{Diagram of the PandaX-I electronics and DAQ system. The red dashed lines
enclose trigger logic components.}
\label{Electronics_and_DAQ}
\end{figure}

A schematic diagram of the electronics and DAQ system
is shown in Fig.~\ref{Electronics_and_DAQ}.
The system is very similar to the one used in the XENON100
experiment~\cite{ref:xenon100}.
The high voltage (HV) mainframe SY1527LC
with four A1932AP 48-channel positive HV modules
were used as the HV supplies for the PMTs in both arrays.
The raw signals from
the PMTs were transmitted
out of the detector together with the HV on the same coaxial cable
and were separated via a capacitive decoupler outside the detector~\cite{ref:pandax_tdr}.
Since the average gain of the PMTs was 2$\times$10$^6$, to achieve a 2$\times$10$^7$
gain as mentioned in Sec.~\ref{sec:requirements},
the raw PMT signals were amplified linearly by a factor of 10 by the Phillips~779
amplifiers before entering into the CAEN V1724 (8 channels per VME module)
digitizers~\cite{ref:V1724_manual}.
The V1274 digitizer stored data in buffers which were read out upon receiving the trigger signal
generated by the trigger system.
The scaler unit (CAEN V830) counted both raw and system trigger signals
to monitor the system deadtime.

\subsection{Digitizers}
\label{sec:v1724}
The central components of the electronics are 23 VME modules
of CAEN V1724 digitizers,
located in two VME crates, connected through optical fibers.
Each V1724 supplies
eight channels of 14-bit 100 MS/s flash ADC (FADC) with 2.25 V$_{pp}$ dynamic range,
leading to 0.137 mV for the least significant bit. Each of the 180
PMT signals is connected to a FADC channel.
Taking the approximate 10~mV~$\times$~20~ns/SPE conversion factor mentioned
in Sec.~\ref{sec:requirements}, the maximum S1 ($\sim$100~PE, 100~ns)
and S2 signals ($\sim$10000~PE, 2~$\mu$s) in the dark matter search region
will have an amplitude of 400~mV and 2~V,
respectively, assuming a triangular shape of the waveform,
even if all photons are collected by a single PMT.


Each V1724 has an independent 100 MHz internal clock.
It allows synchronization among boards
by using the CLK-OUT from a given board as the master, and
daisy-chaining the CLK-IN and CLK-OUT of the rest of the boards with a programmable
phase for each CLK-OUT signal.
All 23 VME boards were verified to be synchronized to better
than 2~ns by routing the CLK-OUT from each board into the oscilloscope.
This is sufficiently precise for this experiment given that a typical S1 signal
is roughly 100~ns wide.

\subsubsection{Majority Signal}
\label{sec:maj}
Each V1724 board houses a 12-bit 100 MHz
DAC with 1 V range on a 50 Ohm load and the output is available on
the ``MON/$\sum$'' connector on the front panel.
Each FADC channel generates a
time-over-threshold signal on a preset threshold (high = 125 mV),
and the internal sum of eight channels is output by the DAC.
The area of this so-called MAJ signal encodes
the overall PE values in this board for small signals when 
PEs are not piling up in the waveform.  
In PandaX-I, the threshold for the majority signal was set to be $\sim$1 PE
for each channel. The MAJ signals are used in the trigger generation, which will be
discussed in detail in Sec.~\ref{sec:trig}.

\subsubsection{Zero Suppression}
\label{sec:zle}
The CAEN V1724 has the capability to record the waveforms in a Zero Length Encoding (ZLE)
mode,
discarding the data under a threshold set by the user.
In PandaX-I, after the gain stage, we chose the amplitude of 1/3 PE
as the threshold for the waveform suppression. The efficiency for such a threshold
was studied by comparing the low intensity LED run with and without the
ZLE to be nearly 100\%. To allow proper baseline
determination, we set the ``look back'' and ``look forward'' samples before and
after the threshold crossings both to 40.
In PandaX physics runs, ZLE mode compresses the data by a factor of 7.
Even with the 30 Hz source calibration data, the achieved data
rate is 30 MB/s, far less compared to the 70 MB/s data acquisition bandwidth
(see Sec.~\ref{ref:daq_performance}).

\subsection{Trigger System}
\label{sec:trig}
The general requirement for the trigger system is to achieve a
low energy threshold $<$1 keV$_{ee}$. To accomplish this, the actual
design has to take into account two important issues.
First, since signals from many PMTs are involved in generating the trigger,
the trigger needs to be robust against coherent noise among channels. In PandaX-I
experiment, we observed coherent
short and fast oscillating pulses
($\sim$100 ns) from electromagnetic pickups occurring at a
frequency of 200 kHz, that originated from the CAEN SY1527LC power 
supply~\footnote{Note that the noise ripple specification for CAEN SY1527LC 
is 30 mV$_{pp}$, and what we observed was within this specification.}.
Under these circumstances, neither number-of-hit
nor summed-signal-amplitude work well as a trigger source.
Second, for the few-PE level S1 signals, it is very difficult to discriminate
against noise in the trigger hardware. In this case, triggers on S2 signals (typically one
hundred times larger) are much more effective. However, for small S2
signals, each PMT receives a few PEs but distributed broadly in a few $\mu$s, a challenge
for the trigger system to identify.

\begin{figure*}[!htbp]
\centering
\includegraphics[width=0.8\textwidth]{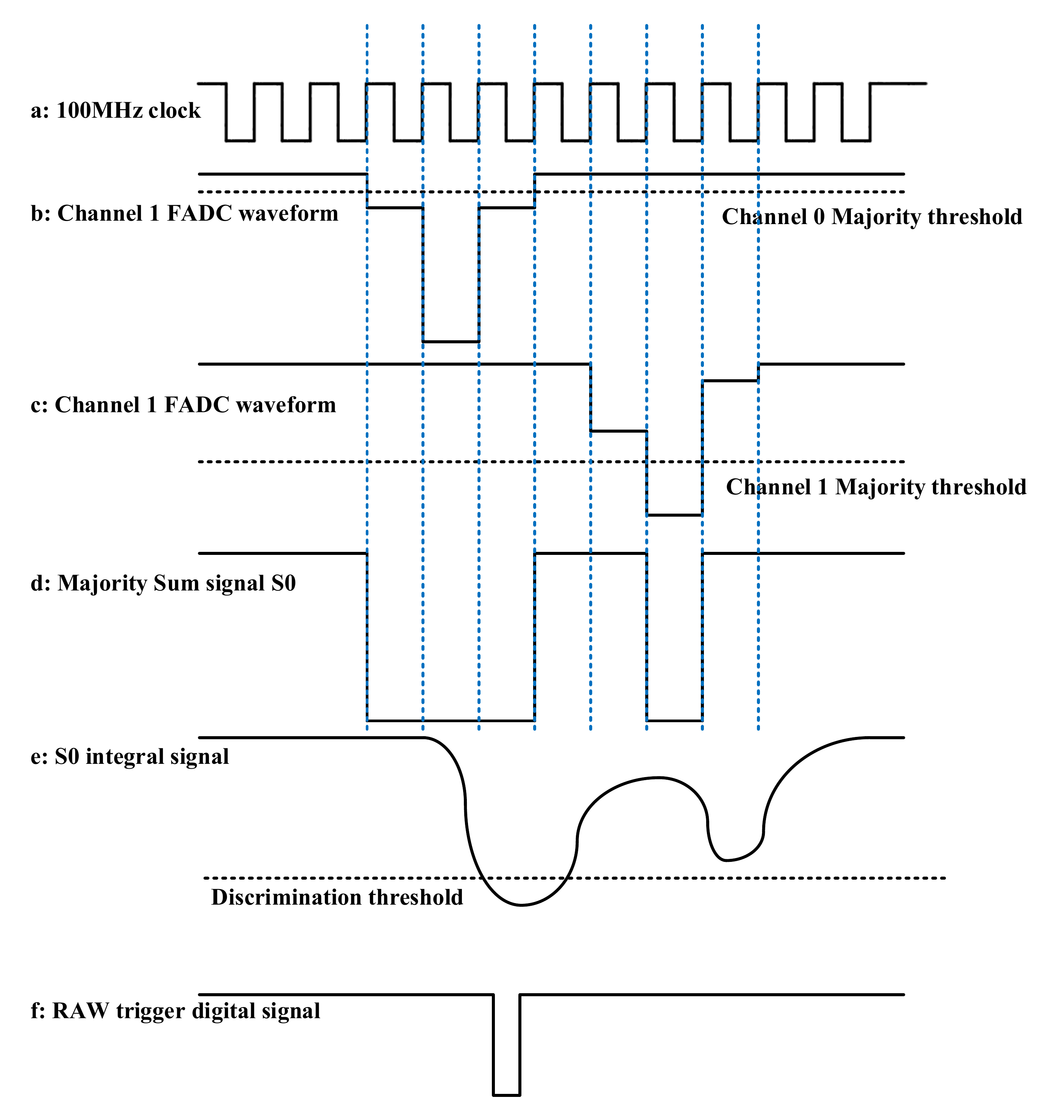}
\caption{a) the 100 MHz clock signal,
b) and c) the waveforms of two FADC channels with dashed line indicating the majority threshold
where the over-threshold samples produce majority output in the unit of 125 mV per channel,
d) the summed majority signal S0, e) the integrated S0 signal after the spectroscopy amplifier,
and f) the digital signal of the raw trigger.}
\label{raw_trigger}
\end{figure*}

In Fig.~\ref{raw_trigger}, an illustration of the PandaX-I trigger generation is shown.
In the PandaX-I detector, both S1 and S2 light collections are dominated by the
bottom 3-inch PMTs, read out by five V1724 modules, each with its own 100 MHz MAJ
output (Sec.~\ref{sec:maj}). As shown
in Fig.~\ref{Electronics_and_DAQ}, a Phillips 740 linear Fan-in/out
module sums up the MAJ outputs from the five modules producing an analogue signal S0,
which is a global time-over-threshold signal with its amplitude representing the
number of fired bottom PMTs. S0 is then integrated by an ORTEC 575A
spectroscopy amplifier
(gain 25, shaping time 1$\mu$s) with the amplitude of the output
proportional to the area of S0.
The integrated signal is further discriminated by
a CAEN V814 VME discriminator to generate the raw trigger signal.
Such a trigger scheme is effective to address the two issues above.
The short coherent noise, even if it happened to cross the majority threshold
on a number of channels, would only produce a narrow S0 signal, whose integral would be
too small to trigger. Small and narrow S1 would not generate the trigger either. On
the other hand, for the broad and low amplitude S2 signals, the integral of S0
approximately represents the overall charge. The discriminator threshold
is set to 15~mV, which is sufficiently low to trigger $<$1~keV$_{ee}$
events efficiently (see Sec.~\ref{sec:elec_perform}).

V1724 uses a circular buffer to store data in its 1~MB memory
to mitigate the deadtime during data readout.
BUSY signals are generated when the buffer is ``almost full''.
The raw trigger
signal and BUSY signals of the 23 V1724 are input into
a CAEN V1495 general logic unit.
To avoid multiple trigger signals within the length of a readout window
(200~$\mu$s), a hold off window of 200~$\mu$s is set after a raw trigger.
The system trigger is generated by the FPGA in V1495 that requires a raw trigger
not being vetoed by any of the BUSY signals.

\section{DAQ System}
\label{sec:daq}
\subsection{Architecture}
\begin{figure*}[!htbp]
\begin{center}
\includegraphics[width=0.7\textwidth]{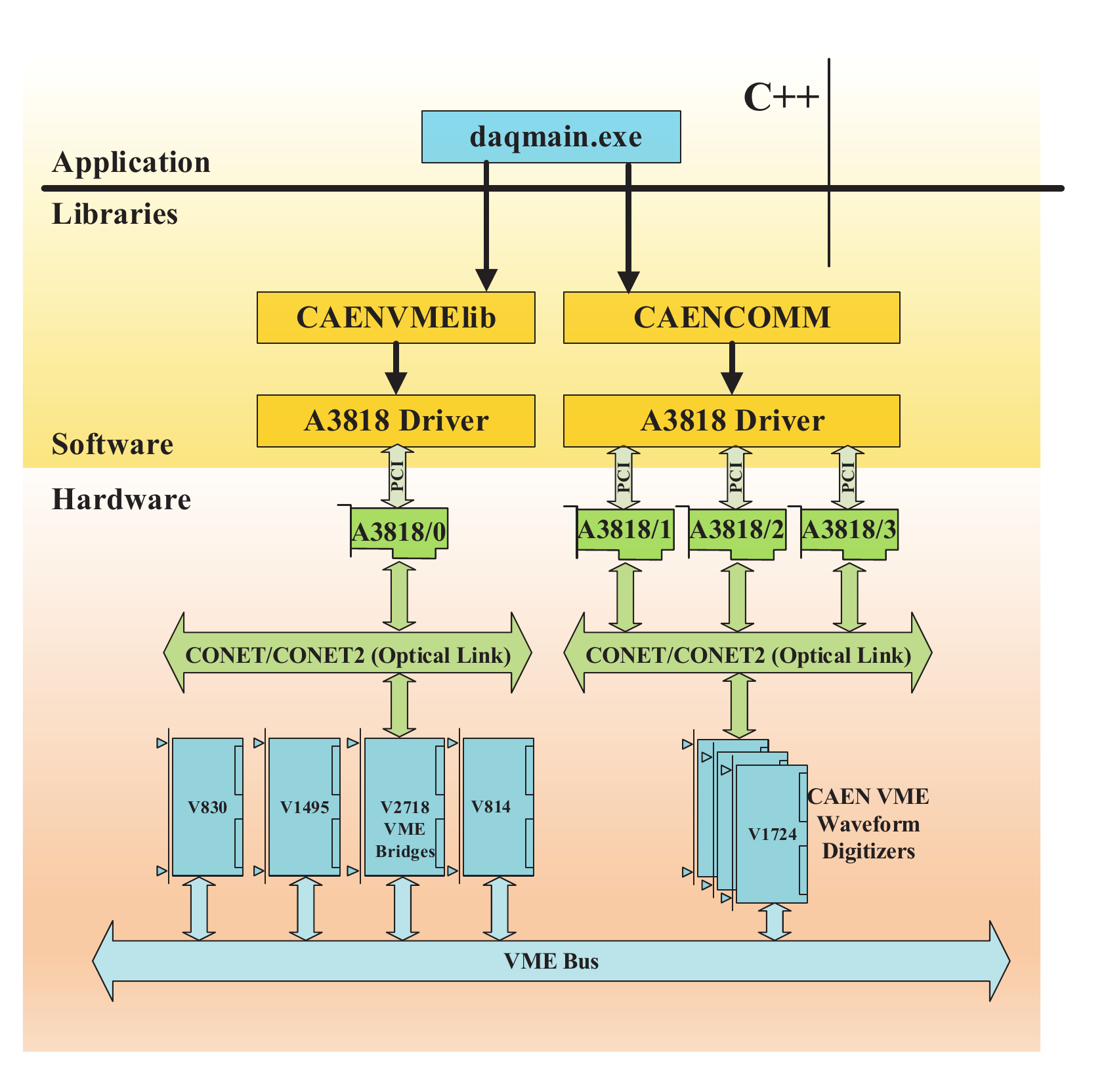}
\end{center}
\caption{Hardware and software architecture of the DAQ system.}
\label{fig:arch}
\end{figure*}

The DAQ system reads out and processes the data from the electronics via
the VME controller and optical bridges.
As shown in Fig.~\ref{fig:arch}, the DAQ server communicates with the VME electronics
through optical fibers with a PCI-e interface card (CAEN A3818C), capable of hosting up to
4 optical fiber links.
The DAQ uses one optical link
to control the VME bus through the controller module CAEN V2718.
The DAQ reads out the data from V1724 directly via three optical links,
each connecting up to 8 daisy-chained V1724.

\subsection{DAQ Software}
The DAQ software was developed in house in C++, with external libraries
from CAEN (Fig.~\ref{fig:arch}), as well as the XML and MySQL libraries.
At the beginning of a run, the system resets all VME modules and clears buffers.
All modules are initialized with configuration
parameters specified in a user XML file.
After the successful initialization, the system accepts the ``run start'' logic, and writes
corresponding run configuration information into a MySQL database.
MBLT (multi block transfer) mode is used to read out every five events stored in the
V1724 circular buffer.

The DAQ contains an event builder to construct the event,
consisting of the header such as DAQ status, configuration, event status,
as well as the FADC data blocks. The FADC data blocks contain auxiliary information about
the VME crate, board number, channel number, and time stamps, etc., to facilitate the
offline analysis.
A long run is broken into number of files, each with a size limit of 1 GB.
At the opening and closing of each data file, the DAQ also inserts
corresponding information into the MySQL database.

An issue was identified in which the data from different V1724s could
occasionally get misaligned, making it nearly impossible to reconstruct the original events.
The issue was due to the loss of synchronization between clocks on different
V1724s. To combat this, the DAQ system used the V1495 module to send out a trigger
time reset signal at the ``RUN START''.
In addition, a method was implemented in the DAQ software to check for a difference
in the timestamps between the first and last V1724 data block within an event,
and forced the run to stop if a problem was found. These implementations fixed
the problem.

In the stable running period, to achieve a complete automation of the DAQ, a background
watchdog program restarted the run if there was a crash.

\subsection{Realtime and Online Data Processing and Monitoring}
\begin{figure*}[!ht]
\begin{center}
\includegraphics[width=0.7\textwidth]{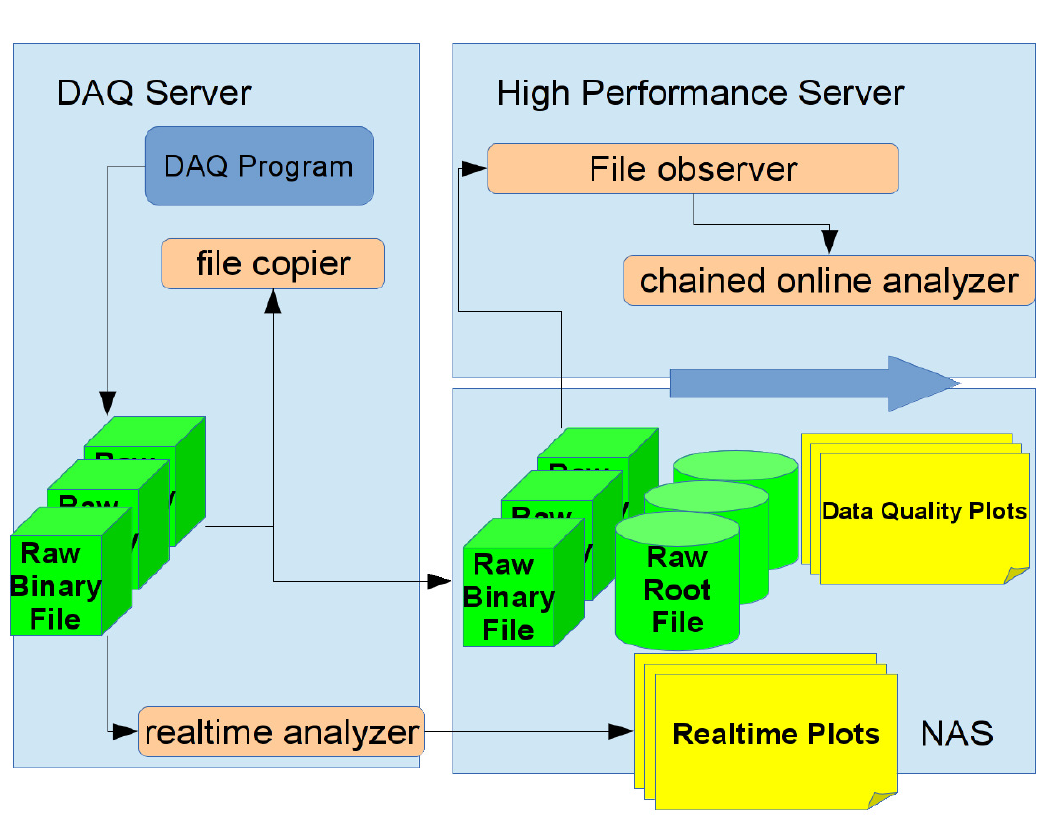}
\end{center}
\caption{The architecture of the realtime and online processing and monitoring system.}
\label{fig:online_monitoring}
\end{figure*}

The online data processing and quality monitoring scheme is shown in
Fig.~\ref{fig:online_monitoring}.
The DAQ server was located underground in the electronics rack with disk arrays
attached to it. To achieve real time monitoring of the DAQ
performance, a process on the DAQ server parsed the output file as the data were
being written onto the disk, producing low-level data-quality plots such as the
trigger rate, PMT hit maps and waveforms, etc., viewable from a web page.
As soon as a file was closed, a file copier program on the DAQ server
would copy it to a network attached storage server
located outside the laboratory, connected
to the underground lab through fiber links. The storage server was attached to a multi-core
processing server. Once the copying of a new file was completed, an analysis program
would be triggered, generating output root files available for online analysis.
Data quality figures of the run, up to the most recent file, would also be
automatically generated, viewable on the web page.

During the smooth running period, the maximum data copying rate was tested to be
60 MB/s, corresponding roughly to a trigger rate of 85 Hz with a
200$\mu$s readout windows on all channels and the baseline suppression enabled~\footnote{Note that the data copying speed depends on the network connection,
which should not be confused with data taking bandwidth between the digitizer and the DAQ server.}.

\section{System Performance}
\label{sec:performance}
The electronics and DAQ system functioned stably during
the PandaX-I data taking. Detailed analysis for the final
dark matter exposure has been published~\cite{ref:pandaxI_final}. Here we
restrict the discussions to the performance of the system.

\subsection{Electronics performance}
\label{sec:elec_perform}
\begin{figure*}[!ht]
\begin{center}
\includegraphics[width=1\textwidth]{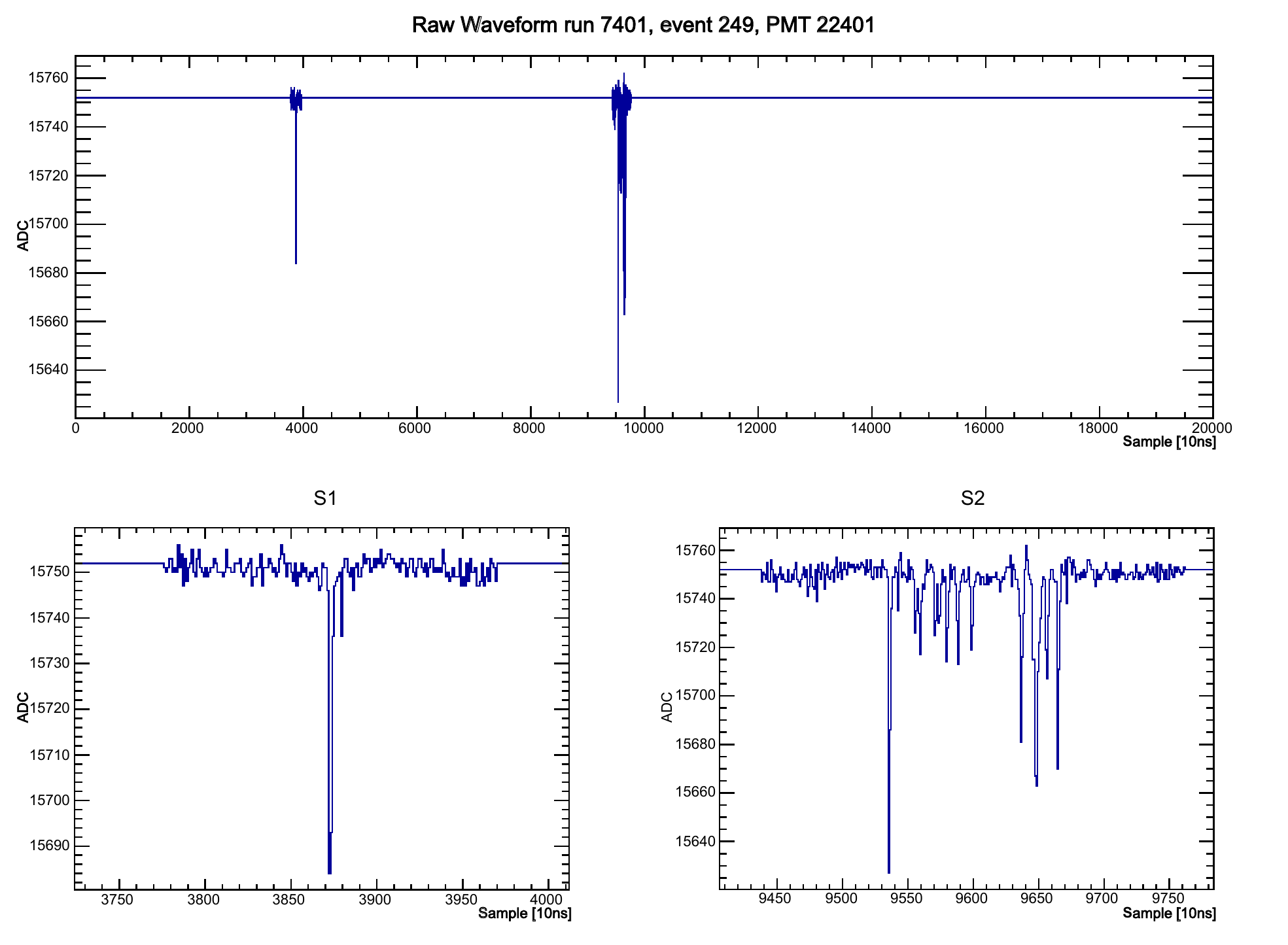}
\end{center}
\caption{A typical PMT waveform from a nuclear recoil event with a single S1 and S2.
  An enlarged view of the S1(S2) waveform is also shown.}
\label{fig:waveform}
\end{figure*}

The central components of the electronics system are the waveform digitizers.
A typical waveform from a 3-inch PMT in a given event
is shown in Fig.~\ref{fig:waveform}, in which
a clear S1 and S2 signal is observed.  The flat regions on the waveform correspond
to the baseline which is suppressed
by the ZLE. For each unsuppressed waveform segment,
sufficient pre- and post-threshold-crossing samples were saved,
based on which the baseline noise could be computed.
On average, the baseline noise was $\pm$ 0.7~mV,
in comparison to the average amplitude of the SPE which was 10~mV.
The noise level obtained during the LED calibration run with ZLE disabled was consistent.

\begin{figure*}[!htbp]
\centering
\subfigure[Charge vs. time for S1s.]
{
\label{fig:ER_qs1_ts1}
\includegraphics[width=0.47\textwidth]{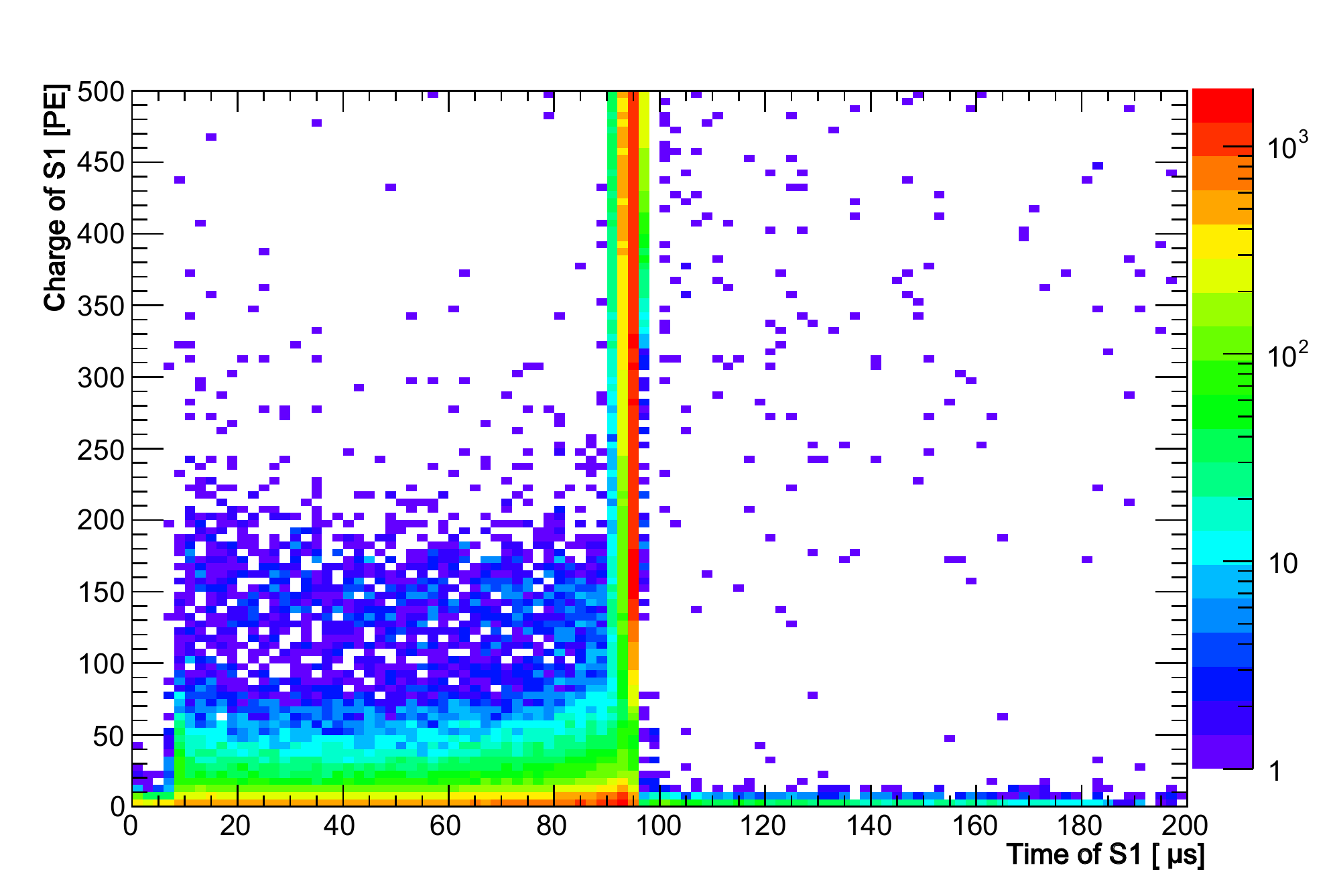}
}
\subfigure[Low energy S2 distribution from the $^{252}$Cf run.]
{
  \label{fig:NR_maj_s2}
  \includegraphics[width=0.47\textwidth]{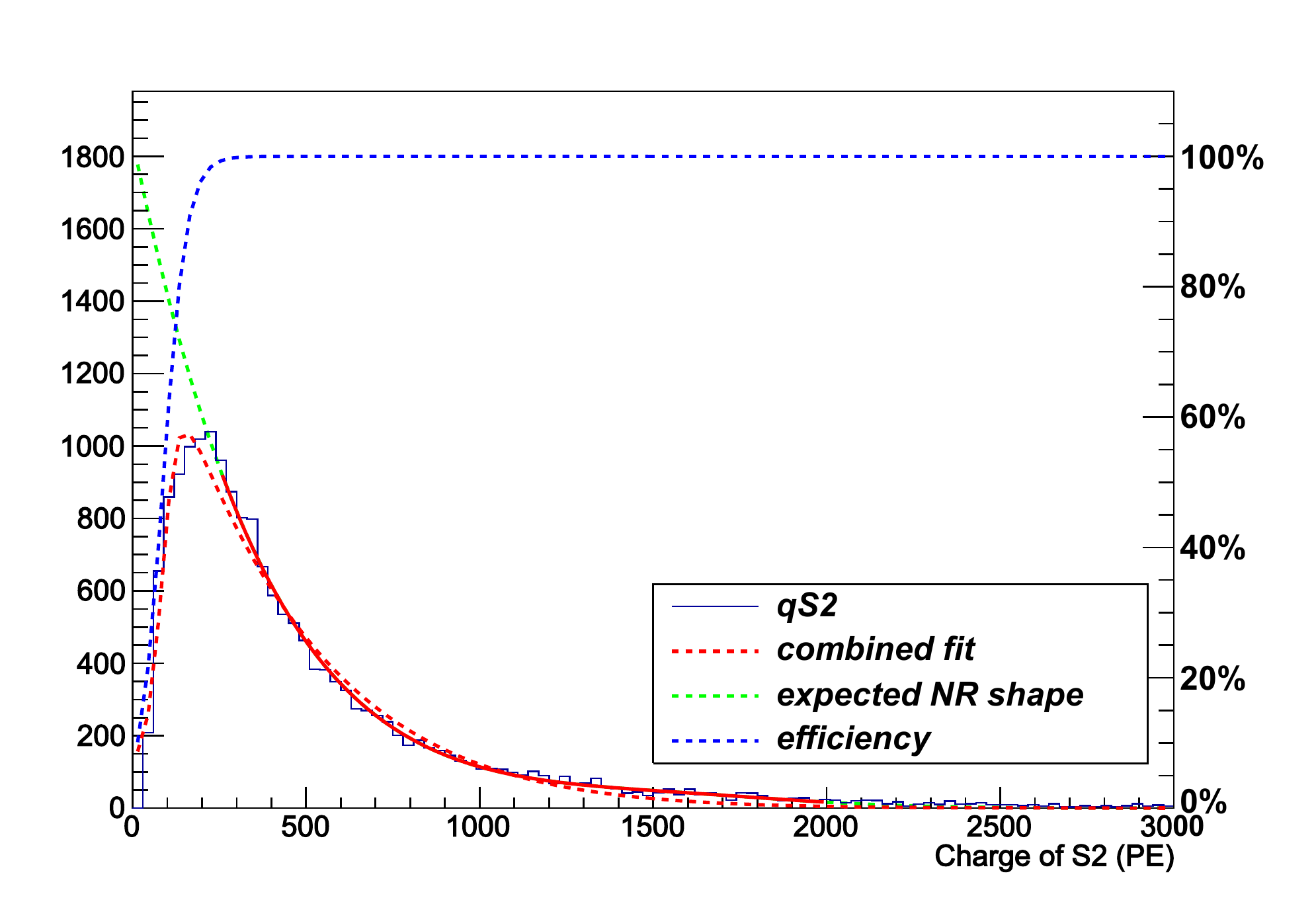}
}
\caption{Distribution of the nuclear recoil events from the $^{252}$Cf calibration run.
a) Charge vs. time for S1s in which the S1 and S2 triggered events can be clearly
identified.
See text for details. b) Fits to obtain trigger threshold for S2 using $^{252}$Cf calibration
data;
green dashed: model for true S2, blue dashed: trigger function (scale on the right), and red dashed: best fit.
}
\label{Trigger Efficiency}
\end{figure*}

The most important performance parameter is the trigger threshold. As mentioned in
Sec.~\ref{sec:trig}, the trigger was generated from the S2
signals for the low energy events in the dark matter search region, and the corresponding
S1s were identified by offline software. In Fig.~\ref{fig:ER_qs1_ts1}, the charge of
S1 vs. the leading edge time of S1 is shown. The bright vertical band around
100 $\mu$s corresponds to the events triggered by S1 with a charge larger than
$\sim$65 PE. The preceding lower band corresponds to the S2 triggered events
with lower values of S1. The events with S1 >100~$\mu$s are from the afterpulsing
of PMTs after large S2 signals but identified as S1s.
To study the threshold
for S2, the low energy
nuclear recoil events in the $^{252}$Cf calibration data were taken with
their S2 distribution shown in Fig.~\ref{fig:NR_maj_s2}.
For these events, the true S2 distribution is expected to approximately be an
exponential shape according to the Monte Carlo. The measured S2 distribution was
then fitted in which the trigger efficiency function was modeled as a Fermi-Dirac function.
The resulting threshold was 89.0$\pm$1.6 PE for a 50\% efficiency.
According to the energy model used in Ref. ~\cite{ref:pandaxI_final},
the mean value of (S1, S2) for a 1~keV$_{ee}$ electron recoil (background)
and nuclear recoil (signal) events are (1.6, 867)~PE and (3.7, 525)~PE,
respectively. Clearly the trigger performance satisfies the threshold
requirement in Table~\ref{tab:req}.

\subsection{DAQ Performance}
\label{ref:daq_performance}
\begin{figure*}[!htbp]
\centering
\subfigure[Data readout rate and live ratio vs. forced trigger rate.]
{
  \label{fig:frequency_vs_bandwidth}
  \includegraphics[width=0.47\textwidth]{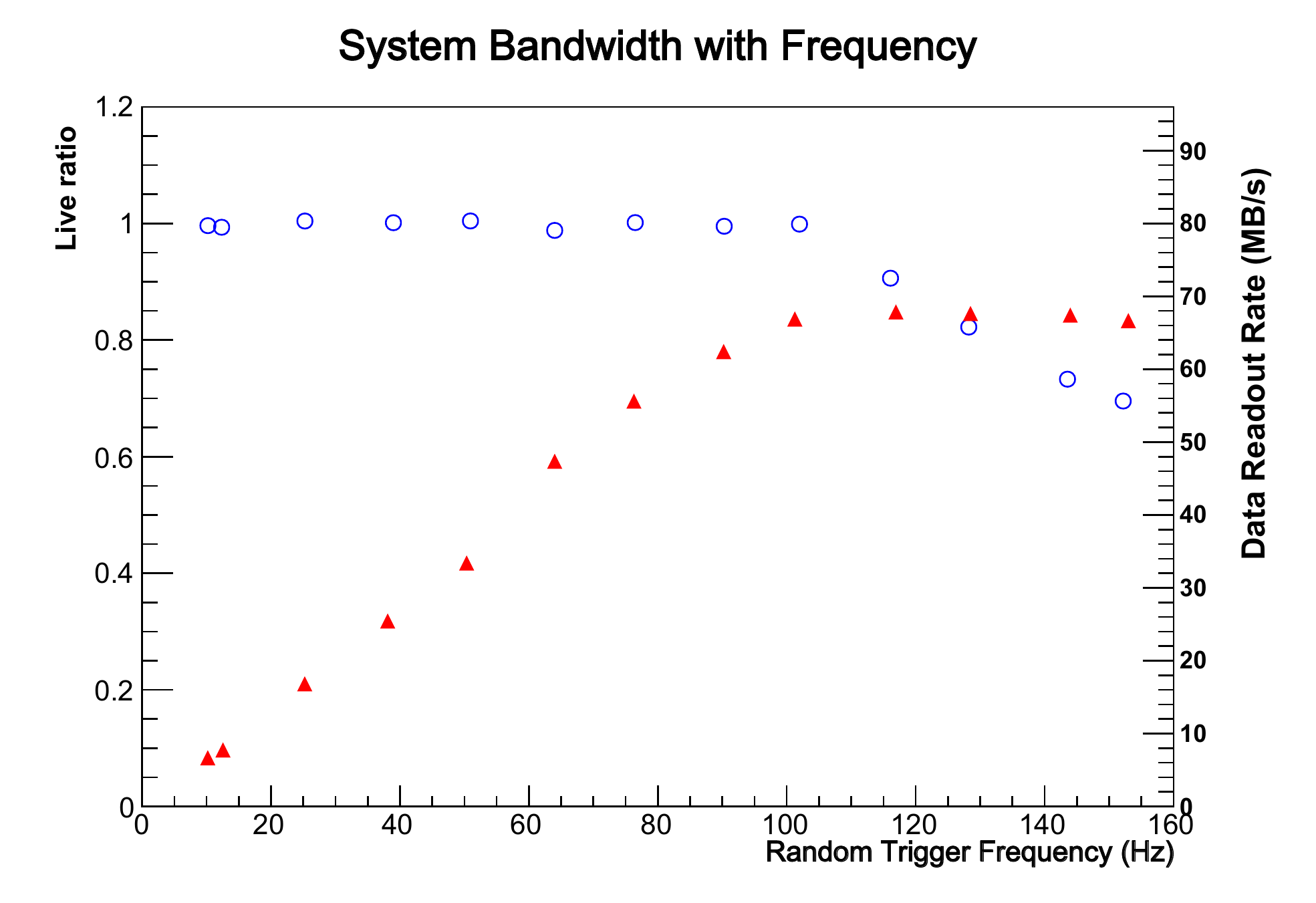}
}
\subfigure[Data readout rate and live ratio vs. event size.]
{
  \label{fig:block_vs_bandwidth}
  \includegraphics[width=0.47\textwidth]{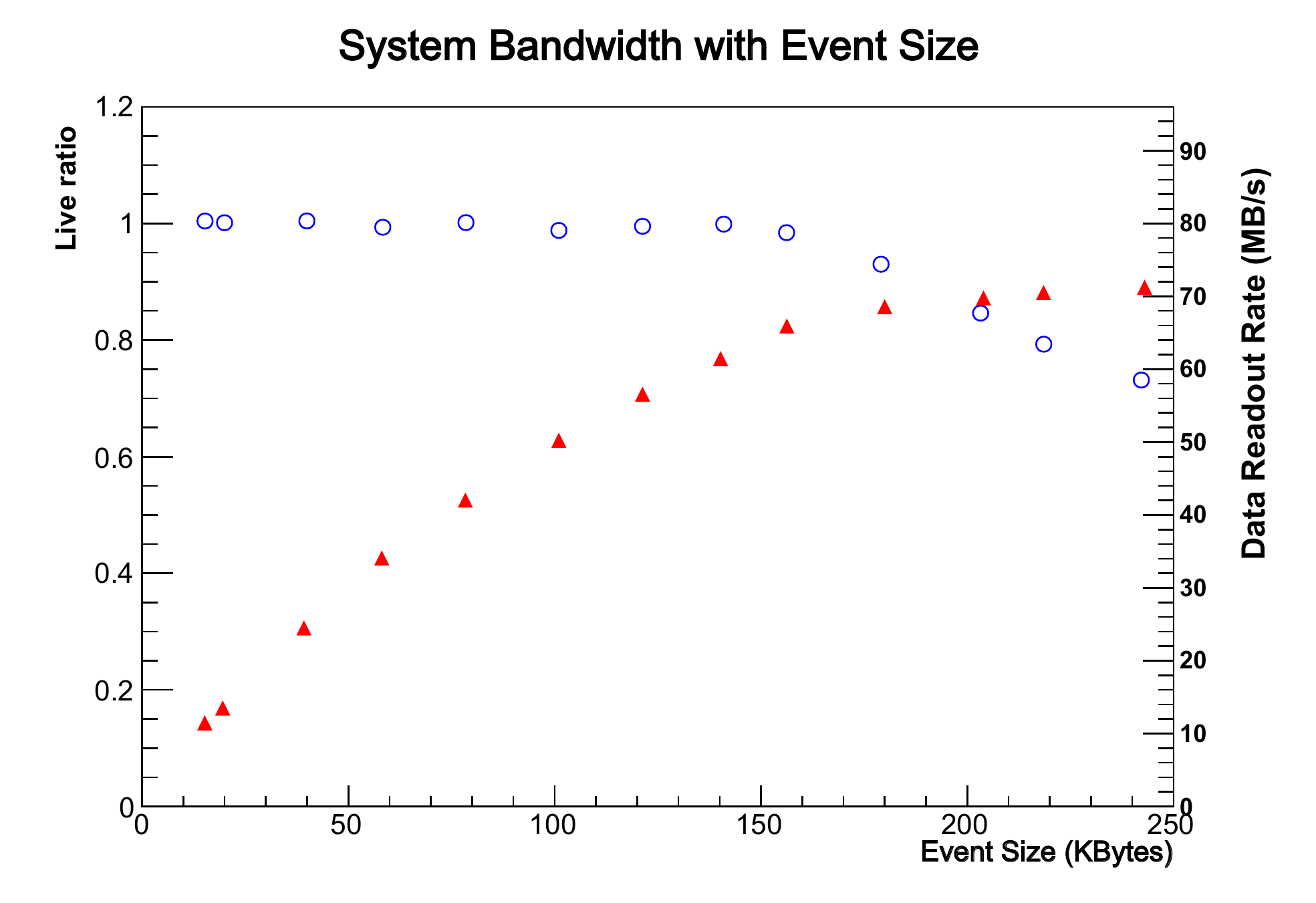}
}
\caption{Studies on the data readout limitation using the LED forced triggers by
  varying the forced trigger rate (left) and event size (right). In both figures,
  red triangles and open circles represent DAQ live ratio (scale on the left) and
  data transfer rate (scale on the right), respectively.
}
\label{fig:Data_readtout_Efficiency}
\end{figure*}

The limitation of the DAQ data rate was measured with the LED calibration
runs without zero suppression by increasing either the rate of the LED
forced triggers, or the length of the readout window. The results
are shown in Fig.~\ref{fig:Data_readtout_Efficiency}.
One sees that in both measurements, the DAQ rate
saturated consistently at around 70 MB/s.

Shown in Table~\ref{performance} are the typical data rates
of the four different run modes in PandaX-I, all
significantly less than the 70 MB/s measured upper limit of the bandwidth.
One expects nearly deadtime-less operation under such rates.
\begin{table}[!htbp]
\centering
\begin{tabular}{c|c|c|c|c}
\hline
Run mode &$^{60}$Co	&$^{252}$Cf	&LED	&DM\\
\hline
Readout window($\mu$s)&   200&   200&   5.12&   200\\
\hline
ZLE&   YES&   YES&   NO&   YES\\
\hline
Frequency(Hz)	&25.32	&22.56	&104.88	&3.58\\
\hline
Event size(kB)	&810.73	&718.61	&224.58	&726.92\\
\hline
Data Transport Rate(MB/s)	&20.05	&15.83	&23.00	&2.34\\
\hline
\end{tabular}
\caption{Basic DAQ parameters and data rate for different run modes in PandaX-I. }
\label{performance}
\end{table}

\section{Summary and Conclusions}
We have described the electronics and data acquisition system for the PandaX-I
dark matter direct detection experiment.
This system used commercial electronics modules but
custom-developed trigger system and DAQ software.
The system ran steadily for more than nine months
and collected about 8 TB of dark matter data. The
performance met the challenging requirements in the search of very low
energy dark matter recoil events.
Although designed for PandaX-I, the same system is used in PandaX-II,
the second phase experiment with a 500-kg target currently under operation ~\cite{ref:pandaxii}.
The increased detector size in fact suppresses the trigger rate due to the self-shielding
effects of liquid xenon,
and the longer drift time does not lead to larger event size due to zero
suppressions in the waveform data.
This system is also expandable to the future phase of the experiment
and maintains flexibility in acquisition strategy to further optimize the performance.

\section{Acknowledgement}
The PandaX project has been supported by a 985-III grant from Shanghai Jiao Tong
University, a
973 grant from Ministry of Science and Technology of China (No. 2010CB833005),
and grants from National Science Foundation of China (Nos. 11055003, 11435008,
11455001, and 11525522).
This work is supported in part by the Shandong University,
Shanghai Key Laboratory for Particle Physics and Cosmology,
Grant No. 15DZ2272100, and the CAS Center for Excellence in Particle Physics.
The work has also been sponsored by Peking University
and the University of Maryland.

\end{document}